%% file: taptype.tex
\documentclass[sigconf]{acmart}

\input{definitions}

\AtBeginDocument{%
  \providecommand\BibTeX{{%
    \normalfont B\kern-0.5em{\scshape i\kern-0.25em b}\kern-0.8em\TeX}}}

\copyrightyear{2022}
\acmYear{2022}
\setcopyright{acmlicensed}\acmConference[CHI '22]{CHI Conference on Human Factors in Computing Systems}{April 29-May 5, 2022}{New Orleans, LA, USA}
\acmBooktitle{CHI Conference on Human Factors in Computing Systems (CHI '22), April 29-May 5, 2022, New Orleans, LA, USA}
\acmPrice{15.00}
\acmDOI{10.1145/3491102.3501878}
\acmISBN{978-1-4503-9157-3/22/04}

\begin{document}

\title[\projname: Ten-finger text entry on everyday surfaces via Bayesian inference]{\projname: Ten-finger text entry on everyday surfaces via\\Bayesian inference}



\author{Paul Streli}
\affiliation{%
 \institution{Department of Computer Science \\ ETH Zürich \country{Switzerland}}
}

\author{Jiaxi Jiang}
\affiliation{%
 \institution{Department of Computer Science \\ ETH Zürich \country{Switzerland}}
}

\author{Andreas Fender}
\affiliation{%
 \institution{Department of Computer Science \\ ETH Zürich \country{Switzerland}}
}

\author{Manuel Meier}
\affiliation{%
 \institution{Department of Computer Science \\ ETH Zürich \country{Switzerland}}
}

\author{Hugo Romat}
\affiliation{%
 \institution{Department of Computer Science \\ ETH Zürich \country{Switzerland}}
}

\author{Christian Holz}
\affiliation{%
 \institution{Department of Computer Science \\ ETH Zürich \country{Switzerland}}
}

\renewcommand{\shortauthors}{Streli et al.}

\input{sections/0-abstract}



\input{figures/_teaser/teaser.tex}

\maketitle

\input{sections/1-introduction}
\input{sections/2-related_work}

\input{sections/3-method}
\input{sections/4-implementation}

\input{sections/5-evaluation-tap}
\input{sections/6-evaluation-text}

\input{sections/7-apps}

\input{sections/8-limitations}

\input{sections/9-conclusion}


\bibliographystyle{ACM-Reference-Format}
\bibliography{TA}


\clearpage
\input{sections/A-appendix}

\end{document}

%% file: definitions.tex
\usepackage{bm}
\usepackage{mathtools}
\usepackage{amsmath}
\usepackage{multirow}

\usepackage{tikz}
\usetikzlibrary{automata,arrows,calc,positioning}

\def\projname{TapType}
\def\optitrack{OptiTrack}

\newif\iftwocolumn
\ifdefined\TWOCOLUMN\twocolumntrue\fi

\newcommand{\paul}[1]{}
\newcommand{\manuel}[1]{}
\newcommand{\ch}[1]{}
\newcommand{\hugo}[1]{}
\newcommand{\andreas}[1]{}
\newcommand{\jiaxi}[1]{}

\newcommand{\TDpaul}[1]{}
\newcommand{\TDmanuel}[1]{}
\newcommand{\TDch}[1]{}
\newcommand{\TDhugo}[1]{}
\newcommand{\TDalice}[1]{}
\newcommand{\TDandreas}[1]{}
\newcommand{\TDjiaxi}[1]{}

\newcommand{\RTDpaul}[2]{}
\newcommand{\RTDmanuel}[2]{}
\newcommand{\RTDch}[2]{}
\newcommand{\RTDandreas}[2]{}
\newcommand{\RTDjiaxi}[2]{}

\DeclareMathOperator*{\argmin}{argmin}
\DeclareMathOperator*{\argmax}{argmax}

%% file: sections/0-abstract.tex
\begin{abstract}

Despite the advent of touchscreens, typing on physical keyboards remains most efficient for entering text, because users can leverage all fingers across a full-size keyboard for convenient typing.
As users increasingly type on the go, text input on mobile and wearable devices has had to compromise on full-size typing.
In this paper, we present TapType, a mobile text entry system for full-size typing on passive surfaces---without an actual keyboard.
From the inertial sensors inside a band on either wrist, TapType decodes and relates surface taps to a traditional QWERTY keyboard layout.
The key novelty of our method is to predict the most likely character sequences by fusing the finger probabilities from our Bayesian neural network classifier with the characters' prior probabilities from an n-gram language model.
In our online evaluation, participants on average typed 19 words per minute with a character error rate of 0.6\% after 30 minutes of training.
Expert typists thereby consistently achieved more than 25\,WPM at a similar error rate.
We demonstrate applications of TapType in mobile use around smartphones and tablets, as a complement to interaction in situated Mixed Reality \emph{outside visual control}, and as an eyes-free mobile text input method using an audio feedback-only interface.

\end{abstract}

%% file: figures/_teaser/teaser.tex
\begin{teaserfigure}
    \centering
    \includegraphics[width=\textwidth]{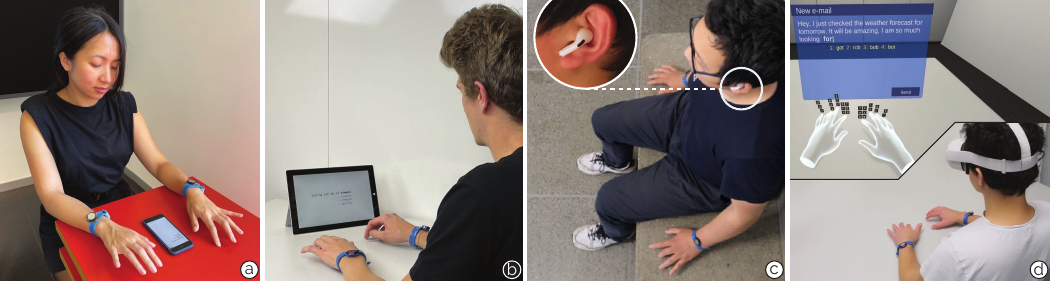}
        \caption{\projname{} is a portable, wireless text entry system that brings touch typing to everyday surfaces.
        TapType's two wristbands sense vibrations arising from finger taps, from which our Bayesian classifier estimates finger probabilities.
        Our text decoder then estimates input character sequences by fusing these predictions with the priors of an n-gram language model given a key-finger mapping.
        TapType is suitable for several applications, including text entry (a)~on a phone or (b)~on a tablet using the surrounding surface for increased typing convenience, (c)~in conjunction with audio feedback only in mobile scenarios, or (d)~in situated Mixed Reality to complement typing with passive haptic feedback.}
    \label{fig:teaser}
    \Description{Figure 1 shows four application scenarios for our text entry system. In the first image, a user is typing on a table to add items to a to-do list on the note app of her smartphone. In the second image, a user is using TapType as a keyboard for his tablet. The entered sentence says 'Typing can be so simple'. In the third image, a user wears earphones and is sitting on a bench. He taps on the bench to enter text, receiving feedback through an audio-only interface. In the fourth picture, the user wears a VR headset and is tapping on the table in front of him to enter text in a virtual e-mail client.}
\end{teaserfigure}

%% file: sections/1-introduction.tex
\section{Introduction}

Physical keyboards have been the go-to option for typing large amounts of text, especially since their commoditization alongside personal computers.
Such keyboards are designed to support fast and bimanual use, while the arrangement of keys allows our hands and fingers to unfold their dexterous capabilities across the full space~\cite{feit2016we, observations136millionkeystrokes}.
A welcome side effect of physical keyboards is their ability to support prolonged use, allowing the user's arms to rest on a surface during continued interaction while providing passive haptic feedback.
This property has allowed them to become a key factor in accomplishing productivity tasks~\cite{hincapie2014consumed}.

With rising popularity of wearable and mobile devices such as smartphones, new types of keyboards have had to compromise on many of these desirable properties in order to improve portability.
Since touch-screen devices integrate input and output into the same surface for \emph{direct} interaction, keyboards necessarily need to fit the available real estate and cannot stretch to full size any longer.
As a result, today's keyboards often appear shrunk~\cite{vertanen2015velocitap} or sparse~\cite{splitboard, Zoomboard}, which affords text input in mobile situations on the go, albeit at the cost of reduced comfort, accuracy, and speed.

To compensate for the input error that comes with smaller layouts that accommodate few fingers, smartphones implement language models to aid in detecting intended keys~\cite{goodman2002language, kristensson2005relaxing, vertanen2015velocitap}.
Researchers have used language models in conjunction with input decoding to port keyboard entry to even smaller surfaces, such as watches~\cite{vertanen2019velociwatch}, and fingertips~\cite{BiTipText, xu2019tiptext}. 
These model-based implementations have since been ported back to soft keyboards on tablets that approach full-size input~\cite{fbhandtracking}, but have also been appropriated to create novel keyboard designs (e.g., One Line Keyboard~\cite{li20111line} or Invisible Typing~\cite{zhu2018typing, shi2018toast}).

In this paper, we introduce \emph{\projname}, a novel text entry system that supports opportunistic, mobile, and full-size touch typing on flat surfaces.
Users simply wear a \projname\ sensor band on either wrist, place their hands down, and start typing.
\projname\ registers taps through inertial sensors and wirelessly offloads events for processing to our backend that predicts entered characters.
\projname's sole requirement on wrist sensors makes it a suitable portable text entry method with the unobtrusive and socially accepted form factor of a fitness tracker.
When typing with \projname, users can transfer their \emph{already practiced and internalized} fast and eyes-free skills of touch typing on a keyboard to any situation on the go.
This makes \projname\ a promising text entry interface for a wide range of mobile, desktop, and spatial-computing use cases.

\subsection{Decoding text entry from accelerometer signals at the wrist}

\autoref{fig:teaser} shows a user entering text by means of our sensor wristband \projname, one strapped to either wrist.
The user types text much like they would on a touchscreen keyboard---arms resting on the surface, using all ten fingers, typing away.
We designed \projname\ to be used for touch typing as taught for text entry on physical full-size keyboards.
This assumption allows us to leverage users' existing muscle memory for surface tapping and to associate pre-defined groups of characters with specific fingers.
Our decoding method then fuses an estimated probability mass function over all fingers for each tap with the priors from an n-gram language model to predict the typed character sequence.

When typing, each tap on the surface causes subtle vibrations that are conducted by the bones in the hand and register with the two inertial measurement units (IMUs) inside the wristband as we showed in our previous project TapID~\cite{meier2021tapid}.
Advancing our previous design, \projname\ picks up on ranging tap intensities, so that tapping requires little intensity, remains non-strenuous, and can thus continue for a prolonged amount of time.
The key component that enables the reliable decoding of taps with \projname\ is our novel Bayesian deep neural network classifier that estimates a better calibrated probability distribution over the fingers following a tap.


Complementing our Bayesian deep network architecture is our probabilistic text entry decoder, which consumes the finger probabilities to output a list of predicted character sequences as suggestions.
Our decoder first relates finger probabilities to groups of alphabet characters according to ten-finger touch typing.
Then it combines the character-specific probabilities with the prior distribution from an n-gram language model.
Our combined approach thus narrows the ambiguous output to a choice of typed words ordered by the estimated likelihood.

For an offline evaluation of our tap classifier and decoder on a challenging dataset, we conducted an experiment where participants typed displayed sentences on a piece of paper on top of a large sensor without receiving any feedback.
Fusing the output of our tap classifier, our decoder pipeline predicts the correct target word in 9 out of 10 cases within the top 5 suggestions.
This compares to only 64\% recall for a tap classifier based on a non-Bayesian neural network of similar complexity as used in TapID~\cite{meier2021tapid}.

In an end-to-end text entry study, we also evaluated \projname\ online with 10 participants. 
Using our system, participants on average typed text at a rate of 19\,words per minute and a character error rate (CER) of 0.6\%.
Expert touch typists consistently achieved an entry rate of more than 25\,WPM in our study at a comparably low error rate.
With respect to \projname's word suggestion, participants picked the correct word in 94\% of all cases when selecting a suggestion.
Across all suggestions, the selected word was at the top of the suggestion list 94 out of 100 times, ranked second 5 out of 100 times, and was lower in 1 of 100 cases.

Finally, we present examples of \projname's potential use in three application scenarios as shown in \autoref{fig:teaser}.
First, \projname's wireless operation affords its use as an extension of mobile devices such as smartphones and tablets as shown in \autoref{fig:teaser}a\&b, respectively. 
In both cases, the user has placed the touch device on the table in front of them to type text, using the surface around or in front of the device for more convenient text input while resting their arms, which also frees screen estate for displaying additional content.
Second, we developed an audio feedback-only interface for \projname\ that enables text input on the go (\autoref{fig:teaser}c).
Here, a user quickly responds to incoming messages with \projname\ by coopting adjacent surfaces.
Our audio component thereby reads out partial phrases, word suggestions, and complete sentences, which makes it suitable for quick message responses on the go.
Third, \projname\ can also address a central challenge in situated mixed reality scenarios.
By opportunistically appropriating passive surfaces, users can leverage a full-size area for typing as shown in \autoref{fig:teaser}d.

\subsection{Contributions}

Collectively, we make four main contributions in this paper:

\begin{itemize}

  \item a novel wristband-based text entry system that enables full-size ten-finger typing on flat surfaces. Our method only takes the microvibrations of surface taps captured by accelerometer sensors inside portable and wireless wristbands as input.

  \item a Bayesian neural network architecture to decode touch events and derive a reliable probability distribution over the fingers of the hand. We evaluated our network in an offline experiment on data gathered from 10 participants who typed on a printed keyboard layout atop a touch sensor, delivering taps unprompted and thus with varying intensity.

  \item a language decoder that fuses the character likelihoods derived from the finger probabilities with the estimated likelihoods for a character from a language model.

  \item an online study where 10 participants typed sentences from MacKenzie and Soukoreff's phrase set~\cite{mackenzie2003phrase} wearing \projname\ bands. On average, participants typed at a rate of 19.2\,WPM with a CER of 0.6\% in the third block (fastest participant: 26\,WPM, CER=0.0\%; slowest: 13\,WPM, CER=0.6\%). 
  
\end{itemize}

%% file: sections/2-related_work.tex
\section{Related Work}

\projname\ is related to mobile text entry systems, inertial sensor-based input detection, and Bayesian deep learning.

\subsection{Mobile \& wearable text entry systems}

\subsubsection*{Touchscreen input decoding}

With the arrival of smartphones, designing text entry systems faced several challenges.
Soft keyboards replaced physical keys, which enabled more compact devices yet raised questions on input reliability, accuracy, and speed, as displaying full QWERTY keyboards on mobile touchscreens leads to small key sizes.
\citeauthor{goodman2002language} introduced a speech-recognition inspired decoder that combined a language model with a probabilistic key press model~\cite{goodman2002language}. 
The model estimated individual key likelihood by fitting a multivariate Gaussian distribution to hits over a given key, which alleviates the notion of rigid key boundaries and led to lower input error.
Subsequent approaches include geometric pattern matching~\cite{kristensson2005relaxing} and Gaussian processes~\cite{gaussianprocess}, trained on gathered touch data to directly predict key probabilities before combining it with a language model. 
The latter approach resembles our method, as we directly estimate a probability distribution over all characters for a given input signal.

Following a large diversity of device form factors, researchers explored decoding and entry performance for various touchscreen sizes~\cite{vertanen2015velocitap}.
Optimized for very small screens, VelociWatch facilitated an entry rate of 17\,WPM~\cite{vertanen2019velociwatch}. 
\citeauthor{gordon2016watchwriter} showed gesture input can reach 24\,WPM on smartwatches~\cite{gordon2016watchwriter}. 
Other work has increased typing efficiency on small screens by grouping characters (e.g., a T9-like keyboard that reached 19\,WPM~\cite{qin2018optimal}).
Others have required repeated input to first pre-select and then make a final selection on an enlarged region for each character (e.g., ZShift~\cite{tinyqwerty}, Splitboard~\cite{splitboard}, Zoomboard~\cite{Zoomboard}).

\subsubsection*{Touch-based input (indirect or without a screen)}

Moving the input away from size-constrained touchscreens frees additional real estate for displaying output.
In turn, everyday surfaces of passive objects can be appropriated for input, which provide passive haptic feedback and at the same time grant more space for bimanual (text) input.
For mobile application scenarios, we require technologies that sense and understand touch from the user's point of view.

Due to occlusion and motion artifacts, hand tracking and particularly sensing touch input using RGB cameras remains difficult.
To detect typing events on a surface, ARKB detects collisions between a fiducial marker-tracked surface and colored fingers~\cite{lee2003arkb}. 
\citeauthor{fbhandtracking}'s temporal neural network decoded typing from finger touches on a flat surface using a motion capture system~\cite{fbhandtracking}, reporting a performance that approaches physical keyboards in an offline study.
Apart from the challenges related to reliability, vision-based methods may also raise privacy concerns.

For non-optical input detection, \citeauthor{goldstein1999non} decoded text from a finger input sequence using trigrams~\cite{goldstein1999non}, speculating that this could be implemented with pressure sensors in the future.
Using an inertial measurement unit (IMU) inside a ring, QwertyRing detects typing with the index finger on an imagined keyboard on a flat surface~\cite{QwertyRing}, estimating character likelihoods from finger orientations.
Our previous project TapID~\cite{meier2021tapid} demonstrated touch and finger detection on passive surfaces using accelerometers inside a wristband, which complemented the hand tracker in a VR headset to enable typing under visual control. 
\projname\ substantially advances TapID's recognition model through our Bayesian tap classifier and, in conjunction with our text decoder, provides a text entry system without the need for camera-based tracking or input under visual control in the first place.

Decreasing the size of the input areas can increase the portability of text entry and allow users to type anywhere.
Related efforts have detected finger touches to register character input in various configurations, such as by appropriating one (12\,WPM)~\cite{xu2019tiptext} or both fingertips (23\,WPM)~\cite{BiTipText} as touchpads for eyes-free selection of individual characters or grouped characters~\cite{fingert9}.
PinchType engages all fingers, selecting character groups from finger pinches that correspond to touch typing on a QWERTY keyboard~\cite{pinchtype}.
Complemented by marker-based hand tracking to robustify pinch detection, their system achieved a mean rate of 12.5\,WPM.

\subsubsection*{Mid-air text entry}

Parallel to touch-based text entry, another thread in the related work investigates mid-air typing.
For example, ATK decodes 10-finger typing in mid-air from a LeapMotion sensor placed below~\cite{ATK}.
Vulture is a word-gesture keyboard that tracks the user's hands with a marker-based system~\cite{markussen2014vulture}. 
Other solutions have used the built-in cameras of mixed reality headsets to detect typing, such as \citeauthor{dudley2018fast}'s approach on a Microsoft HoloLens~\cite{dudley2018fast}.
SCURRY is a glove device that controls a cursor on a virtual keyboard using inertial sensing to capture mid-air typing motions~\cite{kim2005new}.
While mid-air input is promising for typing, especially in mixed reality, it is also known to lead to fatigue over longer periods of time~\cite{jang2017modeling}.
To better support input over a prolonged period, \projname\ builds on our previous approach of leveraging passive surfaces~\cite{meier2021tapid}, which not just allows users to support themselves, but offers passive haptic feedback during input~\cite{insko2001passive,cheng2017sparse}.

\subsubsection*{Non-spatial sensor-based text entry}

For mobile text entry, researchers have proposed alternatives to direct text entry with varying sensing modalities. 
WrisText takes input through joystick-like wrist motions that are detected by infrared sensors inside the band, reaching 9.9\,WPM averaged across a five-day study.
TapStrap is a commercial solution that affixes an inertial sensor ring to each finger to detect chord typing on a surface~\cite{TapStrap}.
Apart from their target use-case of chording input, the company has released a video demonstration of QWERTY-like text entry with two TapStrap devices.
Similarly, Telemetring uses inductive telemetry sensing around each finger for tap detection~\cite{telemetring}, and uses chords for text entry.
While chords allow a one-to-one mapping between keys and finger combinations, they require users to learn a new dictionary.
Both systems rely on the direct instrumentation of the fingers resulting in devices that may be obtrusive, noticeable, and may limit certain interactions.
In contrast, \projname\ consists of only two light-weight wristbands, which are widely socially accepted for wearable technology.

\subsubsection*{Touch typing and memorizing text entry}

Previous research has shown that skilled typists internalize a mental model of the relative keyboard layout~\cite{rashid2008relative, shi2018toast, zhu2018typing}.
This allows them to transfer eyes-free typing skills from a physical keyboard to typing on a flat surface~\cite{findlater2011typing}.
\projname\ benefits from this by leveraging the already acquired spatial input memory~\cite{gustafson2011imaginary} and muscle memory.
Since we only consider the identity of the tapping finger and neglect tap positions, our method tolerates varying relative positions of typing and requires no explicit vertical movement of the fingers.


\subsection{Probabilistic language models}

Language models use the statistics inherent in language to improve text entry performance.
\citeauthor{goodman2002language}'s n-gram language model estimated the prior probability $p(s)$ for a given input string $s$ consisting of characters $y_{1} \ldots y_{l}$~\cite{goodman2002language}.
N-gram models approximate this probability under the Markov assumption that a character only depends on the previous $n-1$ letters.
$p(s)$ is then calculated as 
$
    p(s) = \prod_{i=1}^{l} p(y_{i} | y_{1} \ldots y_{i-1}) \approx \prod_{i=1}^{l} p(y_{i} | y_{i-(n-1)} \ldots y_{i-1})
$.

The estimates $p(y_{i} | y_{i-(n-1)} \ldots y_{i-1})$ stem from a count of letter sequence appearances in a training corpus.
In the case of insufficient data, suitable smoothing techniques can improve estimates~\cite{chen1999empirical}.
The same principle applies on a word level to obtain word n-grams, which are widely used for word, phrase, and sentence-based text entry decoders~\cite{vertanen2015velocitap, wordmultiplewordandsentenceinput, phraseflow}.

Word counts can also help to resolve input sequences on ambiguous keyboards where several characters are typed with the same key~\cite{qin2018optimal, fingert9}. 
Words of a dictionary matching an input sequence can be sorted by their frequency~\cite{qin2018optimal, COMPASS}.

Recent deep neural networks have become a promising alternative for keyboard decoding~\cite{ghosh2017neural, xu2018deeptype}.
Transformers have achieved state-of-the-art results on many language modeling tasks~\cite{vaswani2017attention, devlin2018bert, brown2020language}.

\subsection{Bayesian deep learning}

Deep neural networks tend to be overconfident in classification tasks~\cite{jospin2020hands}, resulting in probability estimates that are much higher than the actual likelihood that the input belongs to a given class~\cite{guo2017calibration}.
Bayesian deep learning provides an alternative where the model allows the assignment of an estimated uncertainty to each prediction.
This uncertainty can be divided into \textit{epistemic} uncertainty (i.e., from insufficient training data, manifesting itself as uncertainty in the weights) and \textit{aleatoric} uncertainty (i.e., from unknown information that prevents a deterministic assignment to a single class)~\cite{der2009aleatory}.

Instead of a point estimate, Bayesian neural networks learn a distribution over the weights $p(\bm{w}|\mathcal{D})$ from a dataset $\mathcal{D}$ using Bayesian inference and compute a marginal distribution over the output $p(y\vert \bm{x},\mathcal{D})$ for an input $\bm{x}$~\cite{wilson2020bayesian}:
\begin{equation}
p(y\vert \bm{x},\mathcal{D})=\int p(y\vert \bm{x},\bm{w}) p(\bm{w} \vert \mathcal{D})d\bm{w}
\end{equation}

Calculating the posterior distribution $p(\bm{w} \vert \mathcal{D})$ from the prior belief $p(\bm{w})$ requires calculating intractable integrals, for which approximate inference techniques are used~\cite{kendall2017uncertainties}.
For Bayesian neural networks, these include Markov Chain Monte Carlo (MCMC) methods including the Metropolis-Hasting algorithm~\cite{chib1995understanding}, variational approximation methods such as Bayes by Backprop~\cite{blundell2015weight}, Monte Carlo Dropout, and others (see \citeauthor{jospin2020hands}'s overview~\cite{jospin2020hands}).

%% file: sections/3-method.tex
\section{Proposed input decoding method}

We now introduce our method that enables ten-finger text entry through touch typing on a flat surface, registered by inertial sensors inside the wristbands and fused with characters' prior probabilities from an n-gram language model.
In addition to the ten fingers, our method detects when users tap the surface with the base of their palms, which activate `delete' and `enter' operations. 
Our method rejects all other inputs including spurious motions and events.

\begin{figure}
    \centering
        \begin{tikzpicture}[node distance={8mm}, thick, main/.style = {draw, circle, minimum size=0.6
        cm}] 
            \node[main] (1) {\footnotesize$\bm{y_0}$}; 
            \node[main] (2) [right of=1] {\footnotesize$\bm{y_1}$};
            \node[main] (3) [right of=2] {\footnotesize$\bm{y_{2}}$};
            \node[main] (4) [right of=3] {\footnotesize\ldots};
            \node[main] (5) [right of=4] {\footnotesize$\bm{y_l}$};
            
            \node[main] (6) [below of=2] {\scriptsize$z_1$};
            \node[main] (7) [below of=3] {\scriptsize$z_{2}$};
            \node[main] (8) [below of=4] {\scriptsize\ldots};
            \node[main] (9) [below of=5] {\scriptsize$z_l$};
            
            \node[main] (10) [below of=6] {\footnotesize$\bm{x_1}$};
            \node[main] (11) [below of=7] {\footnotesize$\bm{x_{2}}$};
            \node[main] (12) [below of=8] {\footnotesize\ldots};
            \node[main] (13) [below of=9] {\footnotesize$\bm{x_l}$};
            
            \draw[->] (1) -- (2);
            \draw[->] (2) -- (3);
            \draw[->] (3) -- (4);
            \draw[->] (4) -- (5);
            \draw[->] (2) -- (6);
            \draw[->] (3) -- (7);
            \draw[->] (4) -- (8);
            \draw[->] (5) -- (9);
            \draw[->] (6) -- (10);
            \draw[->] (7) -- (11);
            \draw[->] (8) -- (12);
            \draw[->] (9) -- (13);
        \end{tikzpicture} 
    \caption{Hidden Markov model illustrating dependencies between a character $y_{t}$ typed at time step $t$ and the corresponding finger tap $z_{t}$ that causes the observed vibration signals $\bm{x}_{t}$. The state of the system is described by the character sequence $\bm{y_{t}}$ entered up and including to character $y_{t}$.}
    \label{fig:bayesiannetwork}
    \Description{The figure illustrates a hidden Markov model which shows the dependencies between an entered character $y_{t}$, its corresponding finger tap $z_{t}$, and the vibration signal $\bm{x}_{t}$ that is observed by our system. The character entered at a time step $t$ only depends on the previous text. The identity of the tapping finger is solely determined by the entered character and shapes the observed vibration signal at $t$.
    }
\end{figure}
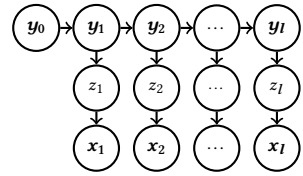

We model our problem in the form of a simple hidden Markov model (\autoref{fig:bayesiannetwork}):
From the accelerometers, we observe the signals $\bm{X_l}=[\bm{x_1}, \bm{x_2} \ldots \bm{x_l}]$ that result from the finger taps $z_{1}, z_{2}, ..., z_{l}$ when typing the letter sequence $\bm{y_l}=[y_{1}, y_{2} \ldots y_{l}]$.
Our goal is to estimate the most likely sequence of characters $\bm{y_l}$ from $\bm{X_l}$,
\begin{equation}
    \argmax_{\bm{y_l}} p(\bm{y_l} \vert \bm{X_l}).
    \label{eq:overallproblem}
\end{equation}

\begin{figure*}[t]
    \centering
    \includegraphics[width=\textwidth]{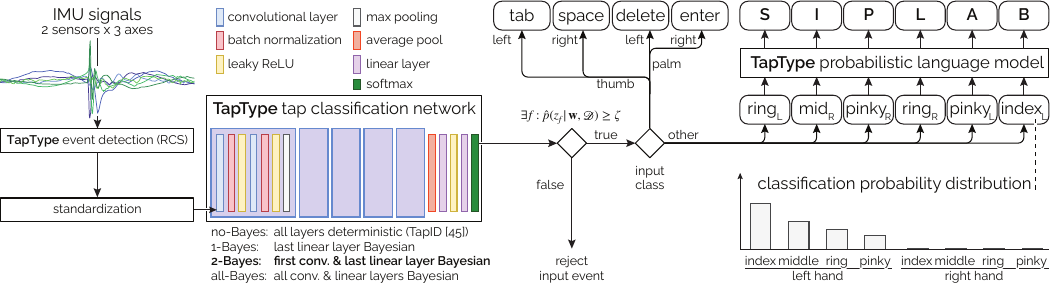}
    \caption{TapType's processing pipeline consists of three parts: 1) a tap detection algorithm identifying sudden changes in the IMU signals, 2) a classification network that estimates the probabilities over the five fingers and the palm of the hand, and 3) a decoder that converts the classifier's output sequence with priors from an n-gram language model to the most likely character sequence.
    	We evaluated several architectures with varying placement of the Bayesian layers on their strength in providing effective probability distributions to the decoder and found \emph{2-Bayes} to produce to highest accuracy and robustness. \RTDpaul{X}{Improve figure.}}
    \label{fig:system}
    \Description{The figure depicts the overall pipeline of our text entry system, which consists of three parts: 1) a tap detection algorithm identifying sudden changes in the IMU signals, 2) a classification network that estimates the probabilities over the five fingers and the palm of the hand, and 3) a decoder that converts the classifier’s output sequence with priors from an n-gram language model to the most likely character sequence.}
\end{figure*}

To implement our method, we introduce the three-part pipeline illustrated in \autoref{fig:system}:
1)~a detection algorithm to register finger taps during typing,
2)~a classification network that outputs a probability distribution over the five fingers as well as the palm of the hand and is also used to discard false positive activations, and
3)~a text decoder that builds on a language model to convert the sequence of probability distributions to a character string.

\subsection{Detecting tap candidates from IMU signals}

To detect a keystroke in the form of a tap, we model each tap as a distinct event within a finite-length window, which is centered on the tap event.
We detect the occurrence of a tap by thresholding the running rate-of-change score $R_{x}$ (adapted from our previous project TapID~\cite{meier2021tapid}).
Our score accumulates the absolute change in the magnitude of the accelerometer signal $\bm{x}^{s}$ corresponding to a sensor $s$ ($\ell_{2}$-norm over the three coordinate axes of a sensor) across time $t$ over all sensors $S$, attenuating past accumulations through the exponential reduction factor $D$,
%
\begin{equation}
R_{x,t} = \tfrac{1}{D}R_{x,t-1} + \Sigma_{s \in S} \lvert{\lVert\bm{x}_{t}^{s}}\rVert_{2}-\lVert\bm{x}_{t-1}^{s}\rVert_{2}\rvert.
\label{eq:RCS}
\end{equation}

Sudden changes in the raw signals cause spikes in the $R_{x}$ function, which indicate contact events. 
When $R_x$ exceeds a threshold at time step $t_{d}$, we determine $t_z = \argmax_{t\in [t_{d},t_{d}+T_{b}]}\,R_{x,t}$ as the time of the tap, where $T_{b}$ is a fixed backoff period.
We then extract a fixed-size window around $t_z$ from all three axes of each sensor.

Note that our method supports a liberal threshold, because it simply serves to reject noisy input.
We reject false activations through spurious events in the subsequent classification step---an event detected at $t_z$ may thus never lead to a character decoding.
This initial threshold thus allows our method to reduce computation and energy dissipation by suppressing processing when idle.

\subsection{Bayesian tap classification \& rejection}
\label{sec:BayesianTapClassifier}

Once a candidate tap event has been identified at $t_z$, we input the resulting window into a classifier network that estimates the probabilities that the tap resulted from each of the fingers or the palm.
These probabilities also aid rejecting spurious activations.

To provide our text entry decoder with more meaningful probability estimates, we adapt a convolutional neural network (CNN) with Bayesian layers to predict the output distribution over the fingers.
This allows our classifier to better convey its \emph{confidence} in a certain prediction.
The notion of confidence is particularly important in the case of an input from a finger that is more likely confused with others (e.g., middle and ring fingers due to similar IMU signatures) or a type of tap that is not well represented in the training set.
In these cases, our Bayesian neural network will distribute the probabilities across several classes, allowing the decoder to consider various options to estimate a character sequence.

For the Bayesian neural network, we employ variational inference where the posterior distribution over the weights $p(\bm{w}|\mathcal{D})$ is approximated by a multivariate normal distribution $q(\bm{w} \vert \bm{\theta})$ with a diagonal covariance matrix. 
Our training procedure minimizes the evidence lower bound (ELBO) corresponding to the Kullback-Leibler (KL) divergence between the two.

Reformulated, this becomes
\begin{equation}
\begin{split}
    \bm{\theta}^{*} &= \argmin\nolimits_{\bm{\theta}} \mathrm{KL}\left[q(\bm{w} \vert \bm{\theta}) \| p(\bm{w})\right] - \mathbb{E}_{q(\bm{w} \vert \bm{\theta})}[\log p(\mathcal{D} \vert \bm{w})]  \\
    &= \argmin\nolimits_{\bm{\theta}} \mathcal{L}_{KL} + \mathcal{L}_{E},
    \label{eq:objective_function}
\end{split}
\end{equation}
where $p(\bm{w})= \mathcal{N}\left(\bm{w};\,\bm{0}, \bm{I}\sigma_{p}^{2} \right)$ is the prior over the weights~\cite{blundell2015weight}.
We then approximate $p(z\vert \bm{w},\mathcal{D})$ by sampling from $\bm{w}^{i} \sim q(\bm{w} \vert \bm{\theta}^{*})$,
\begin{equation}
    \hat{p}(z\vert \bm{w},\mathcal{D})= \frac{1}{N} \sum\nolimits_{i=1}^{N} p(z\vert \bm{x},\bm{w}^{i}).
\end{equation}
The entropy of $\hat{p}(z\vert \bm{w},\mathcal{D})$ can be considered a measure of the total predictive uncertainty~\cite{band2021benchmarking}.
For an out-of-distribution (OOD) tap sample, a Bayesian neural network will likely assign an \emph{almost uniform distribution} over all fingers given the higher uncertainty associated with unseen data.
We further encourage this by training our network on OOD samples that do not belong to tap events by adding an entropic open-set loss term~\cite{dhamija2018reducing}, which enforces the output distribution to be uniformly distributed.

The entropic open-set loss is defined as
\begin{equation*}
\mathcal{L}_{E}(\bm{x})=\left\{\begin{array}{ll}
-\log p(z_{f} \vert \bm{x}, \bm{w}) & \text{if } \bm{x} \text { is from finger } f. \\
- \frac{1}{|\text{fingers}|} \sum\limits_{f \in \text{ fingers}} \log p(z_{f} \vert \bm{x}, \bm{w}) & \text {if } \bm{x} \text{ is OOD sample.}
\end{array}\right.
\label{eq:lossfunction}
\end{equation*}

To reject tap events during runtime, we apply a threshold $\zeta$ to the output of our Bayesian neural network.
This discards inputs that do not have high-enough probabilities assigned to any class as false-positive detections and stops further processing at this point.

To speed up inference using our Bayesian neural network, we employ the local reparameterization trick where we sample the activations instead of each weight individually to enable efficient parallelization~\cite{kingma2015variational, shridhar2019comprehensive}.
Thus, we can infer the outputs for an ensemble of predictions for a tap input with a single minibatch.

\subsubsection*{Comparison with TapID}

While the underlying architecture of our Bayesian neural network classifier may appear similar to TapID's CNN implementation~\cite{meier2021tapid}, the better calibrated output of our classifier is a key \emph{prerequisite} for our text decoder.

TapID's classifier network was trained with a cross-entropy loss, which tends to result in overconfident predictions when the negative log-likelihood loss is extensively minimized on its own.
This leads to a badly calibrated classifier~\cite{guo2017calibration}, where the output of the final softmax function assigns all probability to a single finger and thus only provides information about the most likely class.

For TapID, the classification accuracy was 90\% after fine-tuning on 10 user samples.
This accuracy is impractical for text input, however.
For a five-letter word, say, this method would produce a correct finger sequences in $0.9^5 = 59\%$ of the time.
Such low accuracy would require strong error correction from the text decoder and vastly increase the number of possible matches.


\subsection{Probabilistic text entry decoder}

From the sequence of estimated probability distributions from our Bayesian classifier (one for each tap), we aim to predict the corresponding character sequences and thus the typed words.
For this, we propose a probabilistic text entry decoder that determines the most likely string sequences based on these finger probability \emph{distributions}---instead of considering just the most likely finger for each tap.
We accomplish this by weighing the prior probability that is assigned by a language model for a specific character with the corresponding finger's probability predicted by our classifier:
\begin{equation}
\begin{split}
    & \argmax_{\bm{y_l}} p(\bm{y_l} \vert \bm{X_l}) = \argmax_{\bm{y_l}} \frac{p(\bm{y_l}, \bm{X_l})}{p(\bm{X_l})} =  \argmax_{\bm{y_l}} p(\bm{y_l}, \bm{X_l}) \\
    =~ & \argmax_{\bm{y_l}}  p(\bm{x_l} \vert \bm{y_l},  \bm{X_{l-1}} )  p(y_{l} \vert \bm{y_{l-1}}, \bm{X_{l-1}}) p( \bm{y_{l-1}}, \bm{X_{l-1}})  \\ 
    =~ & \argmax_{\bm{y_l}} \prod\nolimits_{i=1}^{l}  p(\bm{x_i} \vert \bm{y_i},  \bm{X_{i-1}} )  p(y_{i} \vert \bm{y_{i-1}}, \bm{X_{i-1}}).
\end{split}
\end{equation}

In the next step, we apply the conditional independencies following from the hidden Markov model in \autoref{fig:bayesiannetwork}. We make use of the law of total probabilities to rewrite the equation. $p(z_{i,f} \vert y_{i})$ is $1$ for the finger $z_{i,f_{y_i}}$ that $y_{i}$ is typed with and otherwise $0$.

\begin{equation}
\begin{split}
    & \argmax_{\bm{y_l}} \prod\nolimits_{i=1}^{l} p(y_{i} \vert \bm{y_{i-1}}) p(\bm{x_i} \vert y_{i})  \\
    = & \argmax_{\bm{y_l}} \prod\nolimits_{i=1}^{l} p(y_{i} \vert \bm{y_{i-1}}) \sum\nolimits_{f \in \text{ fingers}} p(\bm{x_i} \vert z_{i,f}) p(z_{i,f} \vert y_{i}) \\
    = & \argmax_{\bm{y_l}} \prod\nolimits_{i=1}^{l} p(y_{i} \vert \bm{y_{i-1}}) p(\bm{x_i} \vert z_{i,f_{y_i}}) \\ = 
     & \argmax_{\bm{y_l}} \prod\nolimits_{i=1}^{l} p(y_{i} \vert \bm{y_{i-1}}) \frac{p( z_{i,f_{y_i}} \vert \bm{x_i}) p(\bm{x_i})}{p(z_{i,f_{y_i}})} \\ = 
     & \argmax_{\bm{y_l}} \prod\nolimits_{i=1}^{l} p(y_{i} \vert \bm{y_{i-1}}) p( z_{i,f_{y_i}} \vert \bm{x_i}).
\end{split}
\label{eq:decoderderivation}
\end{equation}

Again, $p(\bm{x_i})$ does not depend on $\bm{y_l}$. 
We estimate $p( z_{i,f_{y_i}} \vert \bm{x_i})$ using the output of our Bayesian classifier. 
In case we train our classifier on a balanced dataset over the fingers, we can assume that $p(z_{i})$ is uniform over all fingers from the perspective of the Bayesian neural network. 
Thus, it does not change with $f_{y_i}$ and can be neglected.
To approximate $p(y_{i} \vert \bm{y_{i-1}})$, we use an n-gram character model. 

Our method finds the most likely character sequence through beam search~\cite{reddy1977speech} and supports our word discovery with an additional word-gram model to improve the final set of results.
Finally, we rank all results according to the sum of the log probabilities from the n-gram character language model, the estimated output of the Bayesian neural network as well as the word-gram model, and present them as suggestions to the user.

%% file: sections/4-implementation.tex
\section{Implementation}

We now describe our specific implementation to arrive at a real-time interactive text entry system.
Our system comprises three components:
1)~the hardware prototypes of the sensor bands,
2)~the design, implementation, and training of our Bayesian classifier, and
3)~our probabilistic decoder with an n-gram language model.

We implemented our system in Python~3.8 to run on an 8-core Intel Core i7-9700K CPU at 3.60\,GHz.
The text decoder ran across multiple cores to parallelize the search, while an NVIDIA GeForce RTX~2080 GPU processed our Bayesian tap classifier.

\subsection{Hardware}

\begin{figure}
    \centering
    \includegraphics[width=\columnwidth]{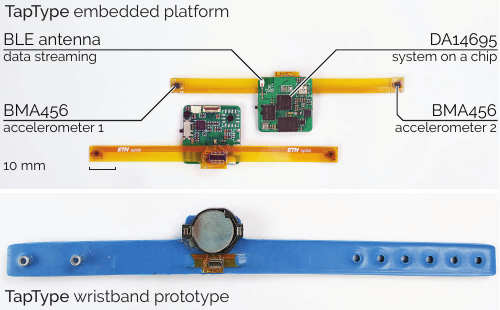}
    \caption{\projname{}'s wristband integrates two accelerometers and a mainboard in a silicone wrist strap~(left). The battery-powered embedded platform~(right) streams the signals via Bluetooth Low Energy to a computer for further processing.}
    \label{fig:hardware_overview}
    \Description{Figure 4 shows a photo of TapType’s wristband which integrates the hardware in a blue silicone wrist strap. Next to it is a photo of the embedded platform. The inertial sensors sit on flex PCB boards that are attached to the mainboard. A system on a chip and a BLE antenna are soldered to the mainboard.}
\end{figure}

For the acquisition of finger and palm tap signals, we developed a wireless sensor wristband as shown in \autoref{fig:hardware_overview}.
With our previous method TapID, we estimated the identity of tapping fingers from the mechanical vibrations recorded with two accelerometer sensors placed close to the ulna and radius.
For \projname, we advanced our previous dual-inertial sensor design with two ultra-low power three-axis accelerometers (BMA456, Bosch Sensortec), which provide higher resolution (16-bit, $\pm$2G) and higher sampling rates (1600\,Hz) for increased precision.
As before, the sensors affix to flex PCB boards that connect to the mainboard as shown in \autoref{fig:hardware_overview}.
A system on a chip (DA14695, ARM Cortex M33, Dialog Semiconductor) downloads the data from the accelerometers and streams it to a backend through Bluetooth Low Energy.
A \projname\ band is powered through a 3\,V CR2032 coin cell battery, which plugs into the holder on top of the board.
\projname\ consumes between 10--15\,mW, lasting for around 30\,hours on a coin cell battery.

For assembly, we cast the electronics in Shore 42 TFC elastic silicone (TFC Troll factory Two-component Silicone Type 15, Riede, Germany) to form a flexible wrist strap that firmly and comfortably attaches to the wearer's skin. 
The silicone also helps couple the tap events to the accelerometers on the flexible straps.

\subsection{Recording training data}

To gather samples for the learning-based approach in our method, we constructed an apparatus to capture representative accelerometer signals for typing input.
Simultaneously, we recorded ground-truth touch events to label accelerometer signals with the correct finger identity.
This setup allowed us to train our classifier in a supervised manner after a collection procedure with participants.

\subsubsection{Apparatus}

As shown in \autoref{fig:datacollection}, our data collection apparatus consisted of two \projname\ bands, a tablet that showed typing instructions to participants, and a sheet of paper with a QWERTY layout printed on it.
Underneath the sheet, a large touch sensor collected input from participants.

\subsubsection*{\projname's IMU streams}

Throughout the data collection, participants wore two \projname\ wristbands, one on either wrist, which continuously streamed the data from the two accelerometers to a PC for logging.
The continuous stream of accelerations included all hand motions including touch and tap events. 
To eliminate the chance that wireless connectivity may drop samples, we connected the wristbands through thin and flexible magnet wires to the PC for power and reliable serial communication.
The tethers ran along the frame and loosely hung from the crossbar to avoid interference during hand motion in the air and on the table.

\subsubsection*{Touch sensor for ground-truth events}

For ground-truth touch events, we use a mutual-capacitance sensor with the dimensions 420\,mm $\times$ 295\,mm (Project Zanzibar sensor~\cite{villar2018zanzibar}).
We printed a QWERTY keyboard layout on an A3-sized paper and attached it atop of the capacitive sensor.
The printed keys had background colors corresponding to ten-finger touch typing on QWERTY keyboards.
The keyboard covered 410\,mm~$\times$~145\,mm and keys measured 29\,mm~$\times$~29\,mm.
We protected the touch area with a sheet of transparent plexiglas (1.5\,mm thick) to prevent wear on the printed keyboard sheet and to solidify our apparatus. 
The digitizer was a Microchip ATMXT2954T2, calibrated to optimally process events through the plexiglas and paper from the copper sensor~\cite{streli2021capcontact}, and configured to output events and coordinates at maximum update rate (> 200\,Hz).

\subsubsection*{Motion capture for 3D fingertip trajectories}

To record the motions of participants' fingertips in midair before and after type events, we additionally mounted four cameras on the frame around the touch apparatus (NaturalPoint OptiTrack Flex13).
Before each recording session, the experimenter attached 2\,mm retroreflective markers to the participant's fingernails.
Because the raw marker positions do not reveal which fingertip a marker belongs to, we recover this information by sorting markers across the forward and right axes and through temporal tracking in the case of a concealed marker.
After the study, we combined the reconstructed fingers with the events and coordinates reported by the touch digitizer to obtain records of finger identities, locations, and all IMU streams.

\begin{figure}[t]
    \centering
    \includegraphics[width=\columnwidth]{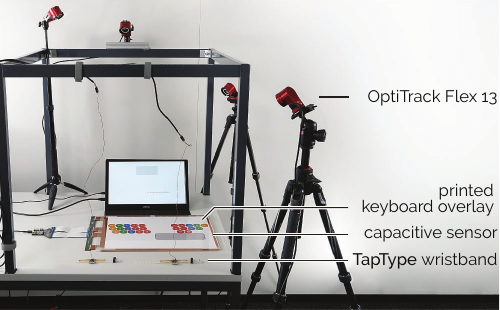}
    \caption{For our data collection, several participants typed sentences on a QWERTY keyboard printed on an A3-sized paper.
    We logged finger tip motions using an \optitrack\ alongside the IMU streams from both \projname\ wristbands.
    For ground-truth touch events and locations, we placed a capacitive touch sensor below the printed keyboard.
    }
     \label{fig:datacollection}
    \Description{The figure shows our apparatus for collecting data. A frame is placed on a table with OptiTrack cameras mounted to it. Below the frame is the paper-printed QWERTY keyboard, which lies on top of a capacitive sensor and is protected by plexiglas.}
\end{figure}

\subsubsection{Participants}

We recruited 10~participants for our data collection (4~female, ages~22--34, mean=28 years).
Participants rated their language and typing skills on a 7-point Likert scale (1 = ``do not agree'', 7 = ``strongly agree'') for the following statements:
``I consider myself a fluent English speaker'' (mean=6.4, min=6),
``I consider myself a fast typist.'' (mean=5.1, min=4, max=7), and 
``My typing style matches the key-finger configuration of this study.'' (mean=4.8, min=2, max=7).
Participants rated the first two statements before the study and gave the third rating afterwards.

We gathered training data with an additional 6~participants (1~female, ages~24--34, mean=27.5 years) from our research group, which we solely used to improve our classifier, but did not include it in any evaluation as test data.

\subsubsection{Task \& procedure}

As shown in \autoref{fig:datacollection}, participants sat at a desk and rested their lower arms on the table.
Participants' first task was to type the text shown on the monitor on the sheet of paper using the respective fingers corresponding to ten-finger touch typing on QWERTY keyboards.
The right thumb activated the space bar.
For each touch, the visual interface acknowledged the key press with the correct character in a Wizard of Oz manner~\cite{zhu2018typing}.
Throughout typing, the experimenter strictly ensured the correct use of fingers with the support of the motion capture software.
When participants had used an incorrect finger for a letter, the experimenter reset the phrase and the participant typed it again.
In a second task, participants produced additional taps at random locations on the surface following visual instructions, completing the capture process for training data.

Overall, our data collection consisted of four blocks and each block had three phases.
In the first phase, participants typed a set of short phrases on the paper-printed keyboard in front of them.
The phrases consisted of pangrams as well as of sentences from Wikipedia that had high entropy over the ten fingers.
This procedure collected at least 15~taps from each finger, except for the left thumb, which was not used for entering text.

In the second phase of each block, the display showed ten squares, spatially arranged to match the fingertip positions of an extended hand.
One after the other, one square was highlighted and participants tapped anywhere on the acrylic glass surface with the corresponding finger for a total of 10~trials per finger and 40~trials for the left thumb.
Highlighted fingers randomly varied across trials. 
After completing the trials for all fingers, participants tapped the palm of each hand on the table a total of 40~times.

In the third phase, participants performed mid-air typing and mid-air movements for another minute with both hands.
This allowed us to collect samples that would cause false activations in \projname's threshold-based activation algorithm, but that could be rejected in the learning-based stage of our pipeline.

Between blocks, participants took off the wristbands, swapped them, and put them on again.
Overall, the data collection took approximately 90~minutes per participant.

Note that participants received no instructions on typing \emph{intensity}.
Because our apparatus reliably detected all touches using the capacitive sensor (and not the rate-of-change threshold), participants' typing events may have been subtle in tapping strength and thus comparable to typing on a tablet.
Therefore, our apparatus allowed us to capture externally valid touch events and intensities.

\subsection{Data processing and candidate detection}

Since all data was logged by the same PC, frames streaming in from the wristbands, the motion-capture system, and the touch digitizer were synchronized on the same clock.
We first processed the fingertip positions from the motion capture to map finger identities to touch events.
For each touch event, we thus obtained the finger identity, touch location, and the typed target character.

We used these touch events to label events in the streams of IMU data.
This also allows us to label all other events in the IMU streams that had no corresponding touch event as spurious input.

\subsection{Bayesian tap classification}

For the design of our classification network, we started with our previous VGG-based CNN to process windows of inertial signals for classification~\cite{meier2021tapid}.
For \projname, we process windows of 128~samples, containing accelerations from both sensors and the three axes each.
These windows therefore contain 80\,ms of input data.

To advance our CNN for Bayesian inference, each layer \emph{could} be converted to a Bayesian counterpart.
Prior work, however, showed superior classification performance for networks with only few Bayesian layers without compromising uncertainty modeling~\cite{zeng2018relevance}.

To compare network architectures, we designed four models.
All models implemented the architecture shown in \autoref{fig:system}: five convolutional blocks that each consist of two convolutional layers with a kernel size of $3\times3$, followed by a max pooling layer.
Each convolutional layer uses batch normalization and a leaky ReLU activation function.
Two linear layers follow the convolutional blocks and a final softmax activation function shapes a distribution over the five fingers and the palm of the hand.
The models differed in the placement of the Bayesian layers.
We substituted Bayesian layers for (a) no layer (\textit{no-Bayes}, e.g., TapID as a baseline~\cite{meier2021tapid}), (b)~only the last linear layer (\textit{1-Bayes}), (c)~the first convolutional layer and the last linear layer (\textit{2-Bayes}), and (d)~all layers (\textit{all-Bayes}).

We trained the models with Bayesian layers using the objective function presented in \autoref{eq:objective_function}. 
For the \emph{non-Bayesian} model, we omit the KL divergence term $\mathcal{L}_{KL}$ in the loss function.
All models produced as output the probability distribution over the five fingers and the palm of one hand having caused an input event.
In case no output is higher than our rejection threshold $\zeta$, the sample is rejected and treated as false positive detection.

Before feeding the data to the classifiers, we standardize the input signals along the time dimension of each axis across the training dataset.
Also, we invert the right axes to mirror the signals from the left wristband across the body's sagittal plane to make use of the inherent symmetry between the human hands.
We then train a single classifier on the samples received from both hands.

\subsection{Text entry decoder}

\projname\ estimates the probability for a given character sequence using an n-gram language model.

\subsubsection*{Training data}
To train our language model, we obtained a dataset of a large number of sentences from Wikipedia, blog posts, news articles, review platforms, e-mails, research papers and other open-source datasets from the internet.
Similar to previous efforts~\cite{vertanen2015velocitap}, we optimized our model for the domain of mobile text entry.
We selected suitable training sentences based on the cross-entropy difference~\cite{moore2010intelligent} between a language model trained on a subset of sentences from a query dataset and an in-domain language model created by combining two separate language models trained on the W3C and TREC 2005 dataset excluding spam messages.
We only used sentences without numeric characters.
Appendix~\ref{sec:appendixLMdata} further details the acquisition and processing of the training data.

\subsubsection*{Language model}

Our implementation of the language models builds on $KenLM$~\cite{heafield-2011-kenlm}.
For characters, we trained a 12-gram character language model with Witten-Bell smoothing~\cite{witten1991zero}.
The vocabulary included all lower case characters \textit{{a--z}} of the Latin alphabet and the apostrophe~'.
For word sequences, we trained a 4-gram word language model with modified Kneser-Ney smoothing~\cite{modifiedknsmoothing} and a vocabulary of 100k English words~\cite{100kvocab}.

\subsubsection*{Decoder}
Our probabilistic text entry decoder receives a sequence of probability distributions over the ten fingers and the palms of the hands as input and a known key-finger mapping that assigns the characters of the vocabulary to the fingers excluding the thumbs.
Given that we know the identity of the hand that has caused the peak and no keys are associated with the thumbs or the palm, the decoder has to consider the probability and corresponding characters of only four fingers for each sample.

Our decoder then sums the log-probability of each finger with the estimated log-probabilities from the character language model for each character corresponding to the respective finger.
To speed up computation, we make use of a beam search algorithm and additionally ignore fingers that have a probability of less than 0.1 assigned by the classifier.
Because users commit a selection after entering a word, we also add the log-probability from the word language model at the end of an input sequence and only return suggestions within our vocabulary.

By design, our current implementation does not account for omission and insertion errors.
However, due to our probabilistic approach, our method supports imprecision in classifying fingers.

\subsection{Interaction vocabulary}

For efficient typing, commands to advance and delete text need to be included, as well as for completing a sentence (`submit') in addition to typing letters.
Because our system is suggestion-based to disambiguate the redundancy of input combinations and compensate for misspelling, another command is required to browse through the list of suggestions.

For entering characters, users type with the four fingers of each hand.
Once our Bayesian classifier reports a typing event, \projname\ plays a click sound to acknowledge each input and, when using visual feedback, appends an asterisk to the already typed characters.
Simultaneously, we feed the sequence of input probability distributions for the currently typed word into our text decoder, which returns a list of suggestions ranked by likelihood.

Our system displays the typed word alongside a list of alternative suggestions once the decoder finishes the processing of the new input sequence, replacing the asterisks with the top entry in the suggestion list.
The user can now accept the suggestion by tapping space (i.e., the right thumb) or cycle through the list of suggestions with the left thumb.
Pressing space at any point accepts the currently selected suggestion.
If the user continues typing with the fingers, the list of suggestions disappears and the word is again substituted with a sequence of asterisks.

To delete the current word, the user taps with the palm. 
Tapping with the left thumb after deleting a word brings up the list of suggestions of the previous word, allowing users to continue cycling through in the cast of an accidental `space'.
Repeated activations of `delete' continue removing words one by one.

To finish a phrase, the user types space twice, similar to the implementation of a full stop on touch keyboards.

\subsubsection*{Probabilistic and deterministic input}

For advancing to the next word, cycling through suggestions, and deleting a word, we chose the thumbs and left palm, respectively, because they produce a distinct input signal in \projname\ that we can reliably classify and thus, process \emph{deterministically}.
Especially for space, the input gesture matches many users' notion of advancing to the next word.
For decoding characters, the uncertainty for input from fingers can be high, making a certain classification difficult and error-prone.
The output of our Bayesian classifier for these fingers is therefore handled by the decoder in a \emph{probabilistic} manner.

\subsubsection*{Out of vocabulary words}

\projname\ supports a mode for entering arbitrary words by selecting characters one by one, even if they are not part of a dictionary.
After typing an individual character, the user can cycle through the list of suggestions with the left thumb and---instead of hitting `space' to advance to the next word---tap with their right palm to accept the suggestion but continue typing the \emph{current} word.
This input process is comparably slow for long words, but expedient for shorter terms, such as abbreviations.

%% file: sections/5-evaluation-tap.tex
\section{Tap evaluation \& optimization}

\RTDpaul{11}{ [recommended but not required] If you have access to the relevant data [or can acquire it], please analyze the difference between users' finger movements when using the suggested typing technique freely compared to the case where the keyboard layout exists.}

In this section, we describe our selection of hyperparameters using the data we collected, including the rate-of-change score threshold, the exponential reduction factor, and the back-off period.
We then describe our validation of the four possible classification networks to determine the best model. 
Finally, we use the results of this evaluation to optimize our pipeline for real-time text entry.

\subsection{Evaluating rate-of-change hyperparameters}

We processed all touch events in our collected dataset and found that the exponential reduction factor $D=1.6$ and a rate-of-change threshold of $R_x=10$ were optimal to detect as many true events as possible while keeping the number of false activations low.
In any case, the latter would be rejected by the classifier.
In conjunction with a back-off period of 64\,samples (i.e., the minimum distance between adjacent peaks), our rate-of-change method detected a touch event in the IMU streams with a recall of 97\% within a 50\,ms window around a tap event registered by our apparatus.

Filtering the dataset with these parameters resulted in 28,116 labeled taps across 16 participants for training.
Processing our collection of mid-air typing and other motions amounted to another 43,529 tap candidates following the rate-of-change score.
The latter candidates represent the OOD samples that we used to train our classifier to reject events.

The hyperparameters for our rate-of-change method face several trade-offs.
Passing more events increases computation load on our network and may risk incurring latency at some point.
We also noticed that samples with lower scores result in a weaker finger classification performance downstream.
Because the capacitive touch sensor detected input events in the study, participants sometimes tapped with little to no force, which still entered the phrase correctly.
These cases, however, lead to minuscule vibrations detected by the sensors, indistinguishable from noise, and without distinctive information about finger identities.

We experimentally determined the minimum rate-of-change score for soft and slow taps at a threshold of 17, which was more than sufficient to pick up taps with reasonable expression.
We used this score in all following experiments.
Using a threshold of 17, 87\% of all recorded touch events are considered.
At the same time, we reduce the amount of false-positive detections by 36\%.

\begin{figure*}
    \centering
    \includegraphics[width=\textwidth]{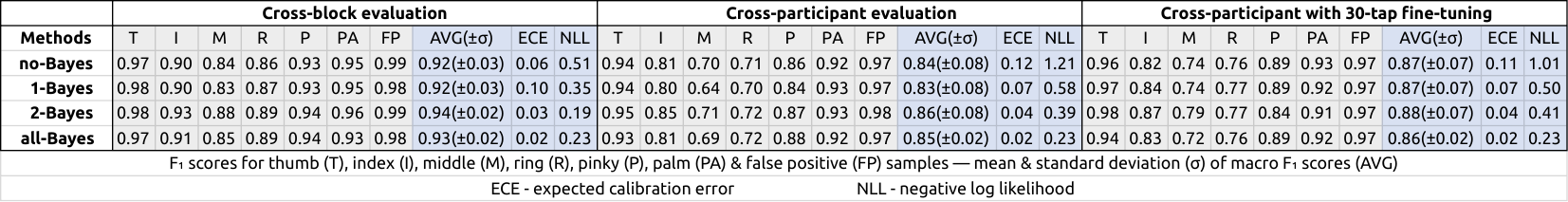}
    \caption{We compared our proposed Bayesian networks with our previous classifier as a baseline (TapID~\cite{meier2021tapid}, labeled \emph{`no-Bayes'}) on the $F_1$ scores, ECE and NLL for cross-session (within-person), cross-person, and cross-person with 30-tap refinement evaluations. We evaluated three network designs: (a)~\emph{1-Bayes} (replacing the last linear layer with a Bayesian linear layer), (b)~\emph{2-Bayes} (replacing the first convolutional layer and last linear layer with Bayesian layers), (c)~all-Bayes (replacing \emph{all} convolutional and linear layers with Bayesian layers).}
    \label{tab:vggcompare}
    \Description{Figure 6 contains a table with the classification results for four different networks. We compare our proposed Bayesian network (2-Bayes) and two other Bayesian classifiers that vary in the placement of the Bayesian layers with the network from TapID (no-Bayes) in terms of the $F_1$ scores, ECE and NLL achieved in a cross-session (within-person), a cross-person, and cross-person with 30-tap refinement evaluation experiment. 2-Bayes achieves the best average performance for all experiments.}
\end{figure*}

\subsection{Classifier validation}

Using the data from the external participants as the test set, we validated the four networks across blocks, across participants, and across participants with per-person fine-tuning.
For cross-block evaluation, we performed four rounds of evaluation per participant, each training on three blocks and validating on the remaining block.
For cross-participant evaluation, we trained the networks on data from 15 participants and validated on the remaining participant for 10 rounds.
For cross-participant with fine-tuning, we took the network trained on $n - 1$ participants, randomly selected 30 taps for each finger from the first block of the remaining participant, and trained with them for another 2000~iterations.
We then evaluated the resulting model on the samples of the other three blocks.

\subsubsection*{Implementation}

We implemented all four networks using PyTorch and trained them for 30~epochs with a batch size of 64 using the Adam optimizer~\cite{kingma2014adam}.
For the Bayesian models, we used an ensemble size of 10 and 128 during training and inference, respectively.
We also balanced the dataset using undersampling.
The OOD-threshold $\zeta$ was set to 0.3 to optimize the rejection accuracy for all four models.
For the isotropic Gaussian prior distribution we use $\sigma_{p}=0.1$.

\subsubsection*{Results}

Table~\ref{tab:vggcompare} shows the classification results for the six classes thumb, index finger, middle finger, ring finger, pinky finger, and palm. 
All four networks achieved similar performance in accuracy.

However, accuracy does not capture the confidence a model assigns to a prediction---yet these confidences are a key enabler for our probabilistic text decoder.
To evaluate the correctness of our classifier to predict the true likelihood of the observed signal to be due to a given finger, we determined the negative log likelihood (NLL) of our model on the test set, which is a lower bound for the evidence~\cite{hastie2009elements}, as well as the expected calibration error (ECE) with 15 bins, which is a commonly used calibration metric~\cite{guo2017calibration, naeiniECE}.
\emph{no-Bayes} now performed significantly worse than the Bayesian networks on these metrics.
While \emph{2-Bayes} was comparatively well calibrated and achieved the highest accuracy, its ECE and NLL were slightly worse than the values for the \emph{all-Bayes} model.

As we are ultimately interested in obtaining the model that most accurately decodes the user input in combination with a trained language model, we also evaluated each model in an offline text entry evaluation in combination with the rest of the decoder pipeline.

\begin{figure*}
    \centering
    \includegraphics[width=\linewidth]{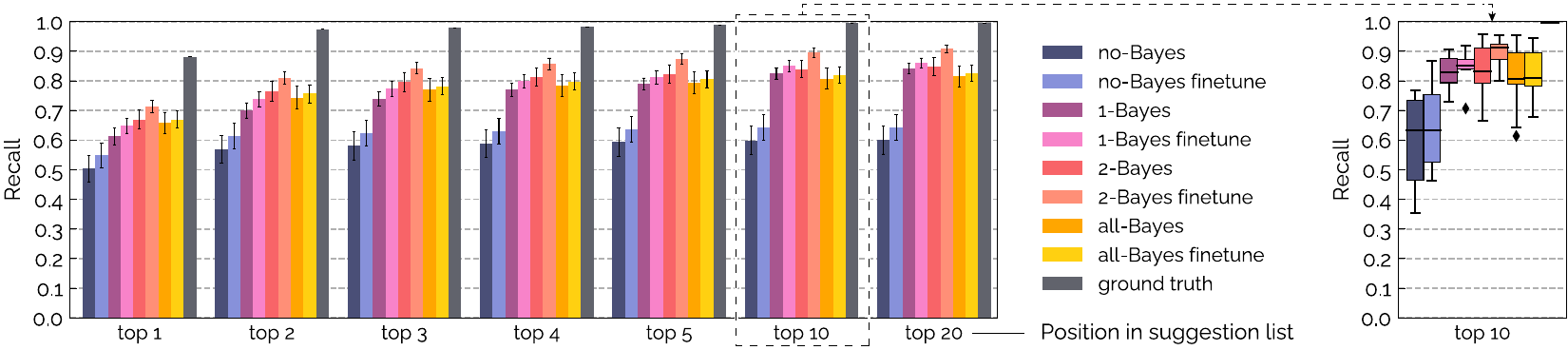}
    \caption{Simulation results of our text entry system by combining different finger classifiers. For each character, we randomly selected a sample of corresponding IMU signals from our dataset and fed them into the finger classifier.
    	We then passed the predicted distribution of finger probabilities into our language model to generate a suggestion list.
    	We counted the number of times the target word occurred in the top 1, 2, 3, 4, 5, 10, and 20 spots, and calculated the respective recall.
    	The chart shows the average recall across participants. Error bars indicate the standard error across participants.
    }
    \label{fig:simulation_results}
    \Description{The figure shows a bar chart visualizing the results of our text entry system by combining different finger classifiers. 2-Bayes with fine-tuning achieves the best performance. Using this classifier, the decoder correctly predicts 9 out of 10 words within the top 10 suggestions. This is a large improvement compared to the no-Bayes network which correctly predicts the target word within the top 10 suggestions in only 2 out of 3 cases.}
\end{figure*}


\subsection{Offline text evaluation for optimization}

To add our text decoder as part of the offline evaluation of our classifier, we simulated text entry on 50 sentences (280 words) from the MacKenzie's phrase set~\cite{mackenzie2003phrase}.
This evaluation also allowed us to systematically optimize the hyperparameters of our system.
Using another cross-participant evaluation (15 training, 1 validation, 10 rounds), we started each round with the models trained on the 15 other participants.
For each letter of the phrase set, we randomly sampled taps from the corresponding key and respective finger in the validation participant's data split, used them as input into the four trained classifiers, which each converted the IMU samples into probability distributions.
In four separate runs, the decoder then processed these outputs to produce a list of word suggestions.
We compared model performances through the positions at which the decoder returned the target word.

\subsubsection*{Results}

\autoref{fig:simulation_results} shows the results of our text entry simulation for the four different architectures.
\emph{No-Bayes} (i.e., the VGG-like network without any Bayesian layers) performed worst, whereas among the Bayesian architectures, \emph{2-Bayes} (i.e., the network with a Bayesian input and output layer) performed best.

\emph{No-Bayes} decoded around 60\% of all words within the top 10 suggestions.
As surmised in Section~\ref{sec:BayesianTapClassifier}, due to the overconfident class predictions, the text entry decoder only considered the characters corresponding to one finger.
This prevented the right word to be proposed as a potential alternative in case of a classification error.

With regard to the performances of the Bayesian neural networks, explaining the differences is challenging.
They all benefit from the improved calibration, with  \emph{2-Bayes} achieving slightly better results than \emph{1-Bayes} and \emph{all-Bayes}.

Notably, \emph{2-Bayes} performs inference about $4\times$ faster than \emph{all-Bayes} (fully Bayesian model).
\emph{2-Bayes} has 2.1~million weight parameters, which is only 0.3\% more than the deterministic \emph{no-Bayes}.

After fine-tuning, \emph{2-Bayes} in conjunction with our text decoder produced the target word in 9 out of 10 suggestions in the top 10 suggestions.
In comparison, a ground-truth classifier for finger identities with 100\% confidence would have proposed 98.9\% of all words within the top 10 suggestions.

%% file: sections/6-evaluation-text.tex
\section{Online text-entry evaluation}

We conducted an online text entry experiment in which participants entered a set of sentences using \projname.
The purpose of this experiment was to determine \projname's efficiency for text entry with touch-typing participants in an end-to-end setting.

\subsection{Study design}

\subsubsection*{Apparatus}

For this evaluation, we kept the apparatus to a minimum to evaluate just our specific implementation.
Participants wore a \projname{} band on either wrist and sat at a table in front of a monitor.
Similar to our data collection, both bands connected to a backend PC through thin magnet wires to remove the impact of potential BLE-based latency, transmission droppage, or power issues.
We reused the frame shown in \autoref{fig:datacollection} to suspend the magnet wires in this study, but we removed all other instrumentation.

For touch detection during the study, we used the hyperparameters established in the previous section.
No capacitive touch sensor aided touch recognition in this study, nor did we include other types of monitoring to obtain ground truth.
\projname\ operated with \emph{2-Bayes} to classify all touch events and perform rejection, trained on the samples from our data collection.

\subsubsection*{Task}

During each trial, participants' task was to enter a sentence through touch typing on the table surface.
The screen in front of participants showed the sentence.
When they produced input through taps, \projname\ behaved as described above, processing the full set of input commands.
Our study application displayed a maximum of ten suggestions, highlighting the first inline as described above.
Participants received no visual support on keyboard layouts or finger mapping during the study.

\subsubsection*{Procedure}

The evaluation comprised eight phases: fine-tuning, keyboard baseline, practice with \projname, four blocks of validation, and a final block of validation without fine-tuning.
Each part contained randomly selected sentences from MacKenzie and Soukoreff's phrase set~\cite{mackenzie2003phrase}, ensuring that no sentence appeared twice in the study or had been used in the previous offline evaluation.
We used the same set of phrases for all participants but randomized their order across participants.
The 100k English word vocabulary for our decoder included all words of the phrase set.
Thus, we added 5 randomly selected sentences from the TWITTEROOV set~\cite{twitteroov}, which contained out-of-vocabulary (OOV) words.
Before the study, participants completed a questionnaire on demographics and on their English and typing skills.
The experimenter also measured participants' wrist circumferences and right-hand lengths.

Next, participants received training for using \projname.
They put on both bands and tapped on the table surface with 30~repetitions with each of their fingers one at a time.
This allowed us to fine-tune our \emph{2-Bayes} classifier.
While fine-tuning the network, participants typed 10~phrases on a physical keyboard with the instruction to be as fast and accurate as possible.
Once the network finished fine-tuning, we deployed the updated weights to our \projname\ backend.
Participants then practiced typing with \projname\ with up to 15 phrases, including the use of gestures for space, delete, and cycle suggestions.
Two of these phrases were randomly picked from TWITTEROOV to practice entering OOV words.

After training, the experimenter started the evaluation.
Across three blocks, participants entered ten sentences per block from MacKenzie's phrase set.
A fourth block with five sentences contained one OOV word per sentence.
Between blocks, participants took 5-minute breaks.
In the eighth and final phase, the experimenter switched the classifier to the \emph{2-Bayes network without fine-tuning} to assess \projname's performance without person-specific calibration.
In this block, participants entered another 10 sentences from MacKenzie's phrase set.

\subsubsection*{Participants}

We recruited 10~participants (2~female, ages 22--40, mean=28.7), accepting only participants who self-described as touch typists.
Wrist circumferences were 150--200\,mm (mean=176\,mm, SD=15\,mm), hands were 170--205\,mm (mean=191\,mm, SD=9\,mm).
Participants received a small gratuity for their time.
In addition, we awarded a remote-controlled quadcopter to the fastest participant with a character error rate below 5\% as an incentive.

On a 7-point Likert scale, all participants rated themselves 6 or higher for ``I am following a system where I consistently hit a specific key with the same finger while typing'' (mean=6.6), 5 and higher for ``I consider myself a fluent English speaker'' (mean=6.3), 4 and higher for ``I consider myself a fast typist'' (mean=5.4), and 5 and higher for ``I consider myself a touch typist'' (mean=6.2).
Before participants judged the latter statement, the experimenter clarified the definition of touch typist and showed the standard 10-finger-system for a QWERTY keyboard.

\subsection{Results}

\begin{figure}
    \centering
    \includegraphics[width=\columnwidth]{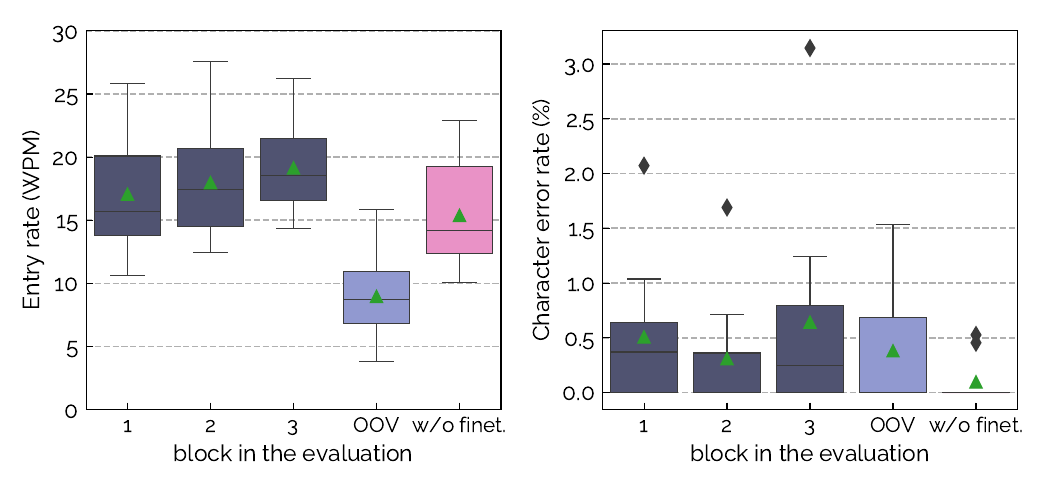}
    \caption{Results for our online text entry study with 10 participants.
    On average, participants entered text at a speed of at least 15\,WPM during the first three blocks (\textit{1}, \textit{2}, \textit{3}), reaching 19\,WPM in the third block with a fine-tuned classifier.
    Text entry rates for phrases with \textit{OOV}-words averaged 9\,WPM.
    Without fine-tuning (\textit{w/o finet.}), participants' speed was around 15\,WPM with a median CER of 0.0\%.}
    \label{fig:results}
    \Description{Figure 8 shows boxplots visualizing the results described in Section 6.2.}
\end{figure}

\subsubsection*{Quantitative performance}

Our main metric to assess performance is text entry rate (words per minute, WPM) and accuracy (character error rate, CER)~\cite{mackenzie_2015}.
We calculated WPM from the time difference between the first tap of a phrase and the selection of the final word for \projname{} and between the first and last entered character for the physical keyboard.
We calculated CER as the Levenshtein distance between predicted and reference text, divided by the length of the reference string~\cite{vertanen2015velocitap}.

On the physical keyboard, participants entered sentences with a mean speed of 74.5\,WPM (min=45.5, max=103.6, SE=6.6) and a CER of~0.4\% (min=0.0, max=1.7, SE=0.2).

Using \projname\ in the three evaluation blocks with the fine-tuned classifier, participants' mean speed in Block~1 was 10.6--25.9\,WPM (mean=17.1, SE=1.5),
12.5--27.6\,WPM in Block~2 (mean=18.0, SE=1.5),
and 14.3--26.3\,WPM in Block~3 (mean=19.2, SE=1.2).
Participants were careful not to make mistakes and quickly corrected errors, resulting in a mean CER of 0.5\% in Block~1 (SE=0.2\%), 0.3\% in Block~2 (SE=0.2\%), and 0.6\% in Block~3 (SE=0.3\%).
We ran a one-way repeated-measures ANOVA with Bonferroni and Greenhouse-Geisser correction to investigate differences in results.
We could not find a significant main effect of Block on entry rate (F(1.17, 10.5)=2.57, p=0.136) or on CER (F(1.44, 12.9)=0.731, p=0.457).
For sentences with OOV words in the fourth evaluation block, where participants individually entered characters by cycling through single letters, mean entry speed dropped to 9.0\,WPM (min=3.9, max=15.9, SE=1.1) at a CER of 0.4\% (SE=0.2\%).
Using our \emph{2-Bayes} classifier without fine-tuning in the final block, participants reached an average speed of 15.4\,WPM (min=10.1, max=22.9, SE=1.4) and a mean CER of 0.1\% (SE=0.1\%).

Participants picked the correct target word in at least 9 out of 10 cases when making a selection.
For the first three blocks with a fine-tuned \emph{2-Bayes}, the target word was the top suggestion in over 90\% of cases and within the top two suggestions in over 98\% of cases. 
For \emph{2-Bayes} without fine-tuning, prediction uncertainty was higher, which is why the recall of the target word within the top 1 suggestions was only 86\%.
However, recall within the top 2 suggestions was over 96\% and over 98\% within the top 3.

\subsubsection{Qualitative feedback}

Participants' general feedback on the use of \projname\ was positive, expressing support for the idea of tap typing.
Several participants with a technical background said that they were a bit surprised that the system worked reliably and some asked whether hidden camera tracking was involved.
Other participants instead expressed ideas for use cases, such as ad-hoc typing in a meeting. 
Several participants also commented that it took them time to trust \projname\ to handle the input, comparing the experience to their experiences with word-gesture keyboards.

\subsection{Discussion}

The results of our online evaluation showed that \projname\ performed well for quick and reliable text entry.
While \projname's entry rates did not approach physical keyboards, participants' average speed of 19\,WPM confirmed our assumption that previously acquired touch typing skills transfer to tap typing on a table.

The live conditions of the evaluation also showed that our model generalized to unseen participants without fine-tuning with the results of the final block.
The lack of fine-tuning accounted for a drop to 15\,WPM (22\% worse) in participants' average speed.

It is worth highlighting that performance considerably varied across participants; one participant typed with over 25\,WPM across all three blocks, reaching rates of up to 44\,WPM on individual phrases.
This indicates that our training session may not have allowed all participants to reach the system limit.
In addition, our impressions observing participants matched their comments on trust in the system and we expect to see an increase in participants' entry confidence and thus speed once they familiarized themselves more with \projname.
We direct the reader to the study footage in our video figure that shows the fastest study participant, who typed confidently throughout all blocks. 

Based on our observations, we interpret the low error rate as participants' intention to make as few mistakes as possible, which led them to correct all erroneous predictions.
This, of course, slowed down the overall text entry rate, but it also demonstrated that \projname\ generally allowed participants to enter the text they intended.

\subsubsection*{Comparison to related approaches}

\projname's average entry rate is 50\% higher than the speed reported for PinchType~\cite{pinchtype}, which also relies on ambiguous ten-finger text entry.
\projname's entry rate is 35\% higher than QwertyRing's 14\,WPM~\cite{QwertyRing}, which participants achieved typing on flat surfaces using IMU ring sensors on the first day.
Using ATK and a displayed keyboard, participants achieved a speed of 29.2\,WPM after four blocks~\cite{ATK}, albeit on a 10k-word vocabulary and using a LeapMotion camera for input.

%% file: sections/7-apps.tex
\section{Scenarios powered by \projname}
We show three scenarios where \projname\ could facilitate eyes-free and ad-hoc text entry.
\autoref{fig:teaser} previews our demonstration apps.

\subsection{Full-size keyboards for smartphones}

In the first scenario, \projname\ complements text entry on smartphones in everyday situations (\autoref{fig:teaser}a).
The user simply places the phone down and types on the table surface next to it, thereby implicitly increasing the input area for the smartphone.
This off-screen typing allows the smartphone to hide the keyboard, presenting more screen real-estate for the running app.
The same benefits hold for tablet interaction, where users can enter text using \projname\ in front of the angled device (\autoref{fig:teaser}b).

We implemented this scenario in React as a web app that communicates with our text entry backend using WebSockets. 
The phone app renders the same feedback as our desktop app, asterisks for words in progress and suggestions laid out above the text field.

\subsection{Audio-only feedback for typing on the go}

Many existing systems implement text entry under visual feedback.
With \projname, we additionally support an audio-only mode that may be particularly beneficial in mobile and ad-hoc settings (\autoref{fig:teaser}c).
Here, \projname\ plays clicks for taps and reads out suggestions as they appear.
When cycling through suggestions, \projname\ reads them out loud and, once the user has selected one, \projname\ reads out the entire phrase.
Similarly, when the user deletes a word, \projname\ plays a delete sound and then reads out the entire phrase to render the system state to the user.
For asynchronous audio playback, we use the \textit{sounddevice} library for Python with playback speed set to $1.5\times$.
Playing a new word stops the previous playback.

\subsection{Text entry in situated Mixed Reality}

Our final scenario is the use of \projname\ in the context of Mixed Reality, here a situated Virtual Reality scenario, where efficient text entry is an important area of research (\autoref{fig:teaser}d).
We continue to take advantage of users' prior experience in touch typing and thus their mental model of text input, enabling a keyboard-free text entry interface in VR.
For guidance, our app can selectively display groups of letters above the user's fingers as an input aid.

\projname's use in spatial user interfaces is not limited to Virtual Reality, but it could equally well work in Augmented Reality scenarios.
Especially productivity scenarios such as office spaces and collaborative tasks could benefit from \projname\ for opportunistic and ad-hoc text entry on passive surfaces such as tables.

We implemented the VR scenario using Unity and combined \projname's events and classifications with the tracking information of the user's hands as provided by the Oculus Quest 2.
Our app renders the user's hands according to the tracker, but triggers input events and key presses based on \projname's pipeline.

We see a particular opportunity for \projname\ to complement input detection of current AR and VR headsets, as \projname\ detects touch and text input outside the headset's field of view.
This enables users of immersive platforms to \emph{indirectly} interact, producing touch input in front of them while looking and focusing their attention somewhere else in the 3D graphical user interface.

%% file: sections/8-limitations.tex
\section{Limitations and Future Work}

\projname's results are promising and indicate that our method is worth exploring further.
In its current implementation, our system has a couple of limitations before it can serve our use-cases in a standalone manner for a wider audience.

\subsubsection*{Mobility}

\projname\ currently relies on neural network inferences for touch classification.
Achieving this in real-time currently requires a high-end CPU or a GPU. 
While we have not focused our efforts on this so far and used a nearby Razer Blade laptop to power all our mobile scenarios, we expect that our method can be implemented on standalone devices, either using emerging neural processing engines or by engineering dedicated model-size reduced classifiers.

\subsubsection*{Wireless transmission latency}
We observed a considerable difference in the transmission latency when the wristbands were connected to the PC with wires compared to wireless BLE transmission.
In our tethered setup, we measured a latency of 59\,ms from the moment of physical contact between finger and surface and the registration of an event by the tap detection algorithm which is similar to the latency observed in our previous project TapID.
In BLE, the latency was 129\,ms over a 1\,m distance.
Latency further increased with growing distances between wristband and receiver.

For processing, our \emph{2-Bayes} classifier incurs an inference latency of 6.5\,ms for a single tap event.
Therefore, after producing an input event, users receive visual or auditory feedback within 65\,ms in a tethered setup, but only within 135\,ms using a wireless connection.
This makes our current implementation feel much more responsive in tethered mode, as tethered latency is smaller than the mean touch duration of $\sim$80\,ms during typing~\cite{kim2013tapboard}.
Before the user is provided with a list of suggestions, the decoder takes around 250\,ms on average to process the classifier's output sequence for a single word. 
In the future, we plan on improving the responsiveness of \projname\ using an improved hardware and antenna design as well as by moving some of our processing on-board.

\subsubsection*{Input ambiguity}

\projname's classifier determines the identity of the finger or palm, but it does not disambiguate between characters that are typed with the same finger.  
This increases the uncertainty in our method and the computational overhead, which makes entering OOV words slow and difficult.
Future improvements of our method could attempt to recover more granular key selections.
In addition to the finger, directly recovering the row or even key would benefit input classification.

\subsubsection*{\projname\ is for touch typing}

Related to input ambiguity, \projname\ relies on participants' ability to perform 10-finger typing.
However, prior research showed that the key-finger mappings vary even among touch typists~\cite{feit2016we}.
Our method already supports changes of the pre-defined mapping through simple reconfiguration without needing to retrain any models.
This makes it suitable for \emph{any 10-finger} typist that \emph{consistently} hits a given key with the same finger.

The formulation in Equation~\eqref{eq:decoderderivation} naturally extends to a probabilistic key-finger mapping where the probability mass of $p(z_{i,f} \vert y_{i})$ is spread across fingers.
This would make the adaptation of our system easier for less-practiced users who do not follow a consistent typing strategy. 
However, the looser prior assumptions would increase the uncertainty of our prediction.
This would then increase latency due to a larger search space and likely decrease the accuracy of our system because of the added ambiguity in plausible solutions.
Given that the key-finger distribution varies across users, we see an opportunity in exploring the use of online learning algorithms for \projname.
These could refine the prior assumptions on a user's typing technique during operation, which would be an interesting direction for future work to improve the efficiency and usability of our method and add a personalization component without overhead on the user's part.

\subsubsection*{Classification performance}

While our \emph{2-Bayes} classifier produced reliable input into our text decoder, it still performed worse than its fine-tuned counterpart. 
To improve our recognition, we plan on expanding our data collection and consider including online learning to constantly improve the classifier when the user selects a suggestion, as is common on smartphone keyboards.
We also expect further improvements with additional correction capabilities of the decoder including insertions, deletions, and substitutions.

%% file: sections/9-conclusion.tex
\section{Conclusion}

We proposed a novel method to decode text from accelerometer signals sensed at the user's wrist using a wearable device.
The key enabler of our method is a text entry decoder that takes the probabilistic output of a Bayesian neural network tap classifier as input and fuses it with the likelihood estimate from an n-gram language model.
We presented \projname, a wireless and wearable smartband text entry device that implements our method and affords text input on rigid surfaces on the go.
\projname\ leverages users' ability to transfer their prior experience from 10-finger touch typing on physical keyboards to \emph{tap typing} on surfaces.
In our online evaluation, participants entered text at an average rate of 19\,WPM, while the subset of more experienced typists achieved an average rate of more than 25\,WPM.
We demonstrated \projname\ potential in applications around mobile text entry to complement smartphones and tablets as well as for ad-hoc situations using audio feedback only.
We see particular promise for \projname\ to support interaction in mixed reality, particularly situated virtual reality, as our approach supports users in entering text \emph{outside visual control}.
Collectively, we believe that \projname\ is a promising enabler for a future generation of wearable and mixed reality systems to type anywhere.

%% file: sections/A-appendix.tex
\appendix 

\section{Language model training data}
\label{sec:appendixLMdata}

The training corpus we used for the n-gram language model comprised sentences from the following datasets:
{\small
\begin{itemize}
    \item  \textbf{20 Newsgroups}: \url{http://qwone.com/~jason/20Newsgroups/} 
    \item  \textbf{Amazon (small)}~\cite{ni2019justifying}: \url{https://nijianmo.github.io/amazon/index.html} 
    \item  \textbf{Arxiv}: \url{https://arxiv.org/help/bulk_data_s3} 
    \item  \textbf{Blog Corpus}~\cite{argamon2007mining}: \url{https://lingcog.blogspot.com/p/datasets.html} 
    \item  \textbf{Enron Email}: \url{https://www.cs.cmu.edu/~./enron/} 
    \item  \textbf{Large Movie Review}~\cite{maas-EtAl:2011:ACL-HLT2011}: \url{http://ai.stanford.edu/~amaas/data/sentiment/} 
    \item  \textbf{Reuters-21578, Distribution 1.0}: \url{https://archive.ics.uci.edu/ml/datasets/Reuters-21578+Text+Categorization+Collection} 
    \item  \textbf{Sentiment140}: \url{http://help.sentiment140.com/for-students} 
    \item  \textbf{TREC2005 Spam (ham)}: \url{https://trec.nist.gov/data/spam.html} 
    \item  \textbf{W3C}~\cite{Craswell-2005-TREC, Benson-2018-found-graph-data}: \url{https://www.cs.cornell.edu/~arb/data/pvc-email-W3C/} 
    \item  \textbf{Wikipedia}:  \url{https://www.tensorflow.org/datasets/catalog/wikipedia}  
    \item  \textbf{Yelp}:  \url{https://www.yelp.com/dataset}  
\end{itemize}
}
We removed sentences with numeric characters and any punctuation except for the apostrophe \textit{'}.

Using cross-entropy difference selection~\cite{moore2010intelligent}, we optimized our training corpus for the domain of mobile text entry.
We randomly picked 10\% of the sentences from the in-domain W3C and TREC 2005 datasets as hold-out data.
As vocabulary for the language models, we took all words that appeared at least twice in the remaining in-domain corpus, which consisted of more than 950,000 sentences in total.
Following \citeauthor{moore2010intelligent}' approach~\cite{moore2010intelligent}, we trained a separate 4-gram language model using back-off absolute discounting~\cite{NEY19941} (implemented with the SRI Language Modeling Toolkit~\cite{stolcke2011srilm} and a discount of 0.7 for all n-gram lengths) on each of the two in-domain datasets.
The models were combined as a mixture model using linear interpolation where the interpolation weights were optimized on the held-out data.

For each non-domain-specific dataset, we trained a language model on a similar number of randomly picked sentences as for the in-domain language models.
We then scored each sentence according to the difference in cross-entropy between the in-domain language model and the model that had trained on the selected sentences from the same dataset.
We trained 4-gram language models on nine different subsets for each dataset.
Each subset only contained sentences with a cross-entropy score higher than a respective threshold that was selected to logarithmically increase the number of samples across subsets. 
For the final training corpus, we only selected the sentences for a given dataset from the subset that achieved the lowest perplexity on the held-out in-domain dataset.
